\documentclass[RNAAS]{aastex62}

\begin{document}

\title{SNIF: The SuperNova Interactive Fitter}
%% Note that the corresponding author command and emails has to come
%% before everything else. Also place all the emails in the \email
%% command instead of using multiple \email calls.
\author{Leilani Baker}
\altaffiliation{Authors contributed equally to this work.}
\affil{Cambridge Rindge and Latin School, 459 Broadway, Cambridge, MA 02138}
\affiliation{Center for Astrophysics \textbar{} Harvard \& Smithsonian, 60 Garden Street, Cambridge, MA 02138-1516, USA}

\author{Sophia Green}
\altaffiliation{Authors contributed equally to this work.}
\affil{Cambridge Rindge and Latin School, 459 Broadway, Cambridge, MA 02138}
\affiliation{Center for Astrophysics \textbar{} Harvard \& Smithsonian, 60 Garden Street, Cambridge, MA 02138-1516, USA}

\author[0000-0002-5814-4061]{V. Ashley Villar}
\affiliation{Center for Astrophysics \textbar{} Harvard \& Smithsonian, 60 Garden Street, Cambridge, MA 02138-1516, USA}

\section{}

Broadband ultraviolet, optical and near-infrared (UVONIR) light curves (LCs) are the most common observational tool used to study supernovae (SNe) and other extragalactic transients. Semi-analytical, one-zone models have been successful in largely capturing the key features of SN LCs and many other transients (see, e.g. \citealt{drout2011first,chatzopoulos2013analytical,nicholl2017magnetar}). Such models are parameterized by only a few parameters: the supernova ejecta mass, the ejecta velocity, the ejecta opacity and parameters specific to the driving energy source (e.g., mass of $^{56}$Ni or initial spin period of a newly born magnetar).

The number of publicly available UVONIR SN LCs has grown exponentially in recent years, especially with the recent creation of the Open Supernova Catalog\footnote{sne.space} which hosts over 14,000 light curves with $>10$ photometric points. However, while the public availability of SN data has increased, open source models and LC fitters remain relatively sparse. Recently, \citet{guillochon2018mosfit} presented the Modular Open-Source Fitter for Transients (MOSFiT), an open-source, Python-based software package which can read in SN LCs from the OSC. MOSFiT (in its default settings) utilizes a Markov Chain Monte Carlo fitter to estimate the best-fit model and model errors. However, this detailed fitting process can take hours to days to fully converge, depending on dataset size.

In this research note we present the SuperNova Interactive Fitter (SNIF\footnote{snif.space}), a tool for fitting UVONIR light curves of supernovae and other extragalactic transients. The interactive software is built on Javascript and Python (Bokeh), utilizing the Open Astronomy Catalogs API (OACAPI, \citealt{guillochon2018open}) to fetch SNe light curves. The light curves can then be quickly fit by hand using parameter sliders. Due its simplicity and accessibility, we hope SNIF will be useful for astronomers, citizen astronomers and students. 

\section{Detailed Description} 

SNIF is a web application built with Python and Javascript. On the associated website, a user can select a supernova model and type in the name of a supernova. If this SN is listed in the OSC, the light curve will be plotted in the available broadband filters.

\begin{figure}[h!]
\begin{center}
\includegraphics[scale=0.45,angle=0]{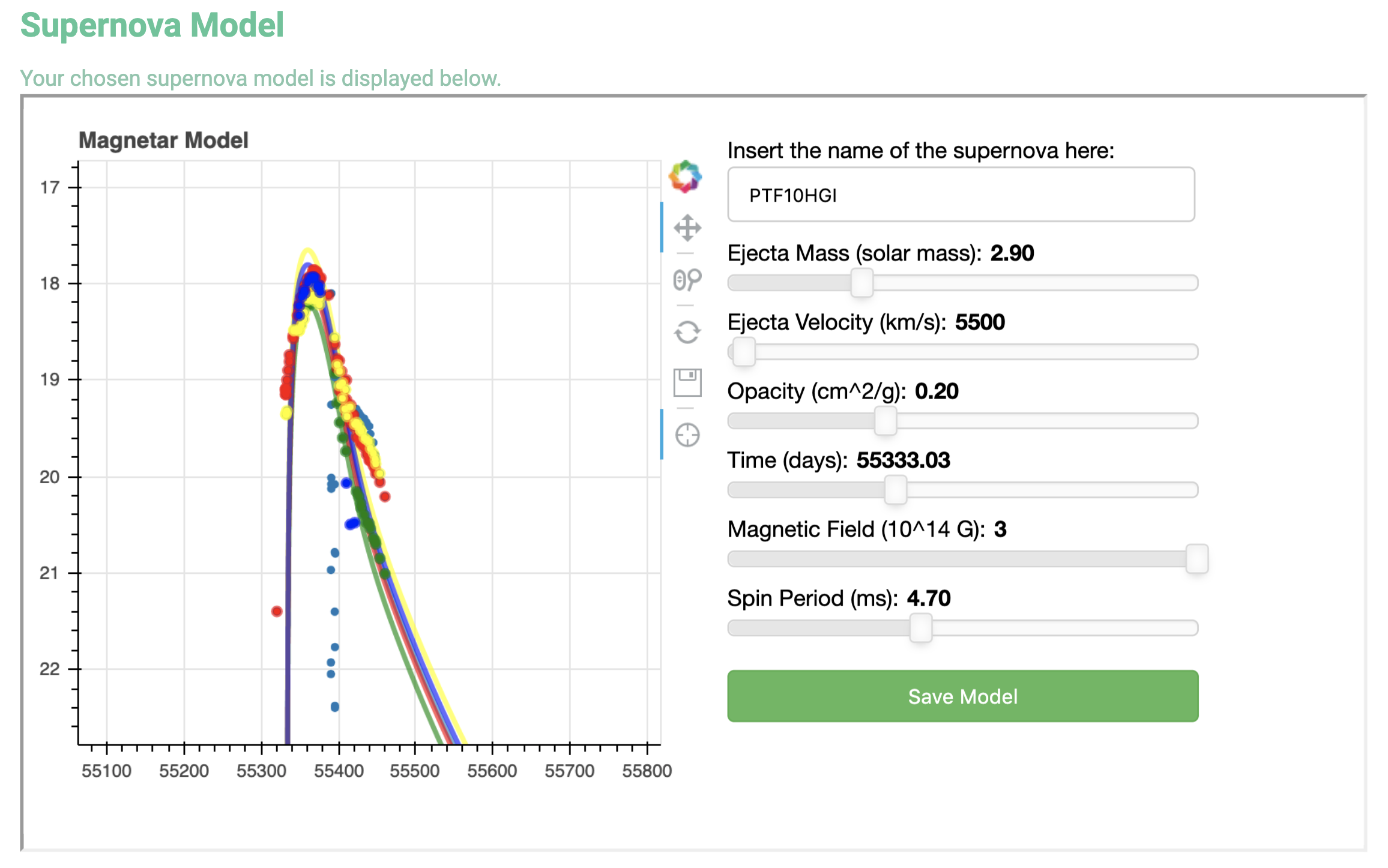}
\caption{Screenshot of an example fit of SN PTF10hgi (Type I superluminous SN) using the magnetar spin-down model. \label{fig:1}}
\end{center}
\end{figure}

Once a SN is selected, the user can use the SNIF's sliders to fit the light curve to a simple analytical, Arnett-like model (see Figure~\ref{fig:1}; \citealt{arnett1982type}). The explosion time, the ejecta velocity, the ejecta mass, and the grey opacity are free parameters in every model. Although the majority of parameter limits are static, the explosion time range will actively change depending on the user's current window size and position. We describe the three currently available models below:

\begin{enumerate}
  \item \textbf{Radioactive $^{56}$Ni-decay:} Radioactive decay of $^{56}$Ni is the most common heating source for supernovae. This model is appropriate for Type Ia and Ibc SNe. The only additional free parameter of this model is the fraction of $^{56}$Ni in the SN ejecta \citep{arnett1982type}.
  \item \textbf{Magnetar spin-down:} The birth and spin-down of a magnetar can heat the inner SN ejecta as the magnetar's angular momentum is converted to kinetic energy. This model is appropriate for Type I superluminous SNe. The additional free parameters of this model include the initial spin period of the magnetar and its initial magnetic field \citep{nicholl2017magnetar}.
  \item \textbf{Ejecta-Circumstellar material interaction:} Shocks induced by the interaction of SN ejecta and pre-existing circumstellar material (CSM) can be a powerful heating source for a variety of transients, including Type IIn SNe and eruptions of massive stars. The additional free parameters of this model include the ejecta profile, the inner CSM radius and the inner CSM density \citep{chatzopoulos2013analytical,villar2017theoretical}. 
\end{enumerate}

When a satisfactory fit is found, the user can save the model parameters by clicking a "Save" button. The JSON save file follows the standard parameter naming schema of {\tt MOSFiT}. The JSON file can be utilized as an initial walker position for {\tt MOSFiT}; an example script is located on Github\footnote{\url{https://github.com/villrv/SRMP_LCs/tree/master/python/example_code}}. 

All source code is publicly available on Github\footnote{\url{https://github.com/villrv/SRMP_LCs}}.

%% An example figure call using \includegraphics

\acknowledgments

The authors are grateful to S. Gomez for helpful discussions throughout this work to and to J. Guillochon for help connecting to the OSC and helpful feedback. LB and SG were participants in the Science Research Mentoring Program at the Center for Astrophysics \textbar{} Harvard \& Smithsonian \citep{2018arXiv180908078G}. Support for this program is provided by the National Science Foundation under award AST-1602595, City of Cambridge, the John G. Wolbach Library, Cambridge Rotary, and generous individuals. VAV is supported by a Ford Foundation Dissertation Fellowship.

%\software{Astropy \citep{http://dx.doi.org/10.1051/0004-6361/201322068}, Bokeh \citep{bokehcitation}, Matplotlib \citep{http://dx.doi.org/10.1109/MCSE.2007.55}, MOSFiT \citep{https://doi.org/10.5281/zenodo.322631}, SciPy \citep{https://doi.org/10.5281/zenodo.3364158}}

\end{document}